\documentclass[manuscript,sigconf]{acmart}

\AtBeginDocument{%
  \providecommand\BibTeX{{%
    \normalfont B\kern-0.5em{\scshape i\kern-0.25em b}\kern-0.8em\TeX}}}

\usepackage{tabularx}
\usepackage{soul}
\usepackage{cleveref}
\usepackage{setspace}
\DeclareCaptionType{vignette}[Vignette][List of Vignettes] 

\renewenvironment{quote}
  {\vspace{2mm}\list{}{\rightmargin=.5cm \leftmargin=.5cm}%
   \item\relax}
  {\endlist}

\copyrightyear{2024}
\acmYear{2024}
\setcopyright{rightsretained}
\acmConference[CHI '24]{Proceedings of the CHI Conference on Human Factors in Computing Systems}{May 11--16, 2024}{Honolulu, HI, USA}
\acmBooktitle{Proceedings of the CHI Conference on Human Factors in Computing Systems (CHI '24), May 11--16, 2024, Honolulu, HI, USA}
\acmDOI{10.1145/3613904.3642115}
\acmISBN{979-8-4007-0330-0/24/05}

\begin{document}

\title{Designing Multispecies Worlds for Robots, Cats, and Humans}

\author{Eike Schneiders}
\affiliation{%
  \institution{University of Nottingham}
  \city{Nottingham}
  \country{United Kingdom}}
\email{eike.schneiders@nottingham.ac.uk}

\author{Steve Benford}
\affiliation{%
  \institution{University of Nottingham}
  \city{Nottingham}
  \country{United Kingdom}}
\email{steve.benford@nottingham.ac.uk}

\author{Alan Chamberlain}
\affiliation{%
  \institution{University of Nottingham}
  \city{Nottingham}
  \country{United Kingdom}}
\email{alan.chamberlain@nottingham.ac.uk}

\author{Clara Mancini}
\affiliation{%
  \institution{The Open University}
  \city{Milton Keynes}
  \country{United Kingdom}}
\email{clara.mancini@open.ac.uk}

\author{Simon Castle-Green}
\affiliation{%
  \institution{University of Nottingham}
  \city{Nottingham}
  \country{United Kingdom}}
\email{simon.castle-green@nottingham.ac.uk}

\author{Victor Ngo}
\affiliation{%
  \institution{University of Nottingham}
  \city{Nottingham}
  \country{United Kingdom}}
\email{victor.ngo@nottingham.ac.uk}

\author{Ju Row Farr}
\author{Matt Adams}
\author{Nick Tandavanitj}
\affiliation{%
  \institution{Blast Theory}
  \city{Brighton}
  \country{United Kingdom}}
\email{{ju,matt,nick}@blasttheory.co.uk}



\author{Joel Fischer}
\affiliation{%
  \institution{University of Nottingham}
  \city{Nottingham}
  \country{United Kingdom}}
\email{joel.fischer@nottingham.ac.uk}

\renewcommand{\shortauthors}{Schneiders et al.}

\begin{abstract}

We reflect on the design of a multispecies world centred around a bespoke enclosure in which three cats and a robot arm coexist for six hours a day during a twelve-day installation as part of an artist-led project. In this paper, we present the project's design process, encompassing various interconnected components, including the cats, the robot and its autonomous systems, the custom end-effectors and robot attachments, the diverse roles of the humans-in-the-loop, and the custom-designed enclosure. Subsequently, we provide a detailed account of key moments during the deployment and discuss the design implications for future multispecies systems. Specifically, we argue that designing the technology and its interactions is not sufficient, but that it is equally important to consider the design of the `world' in which the technology operates. Finally, we highlight the necessity of human involvement in areas such as breakdown recovery, animal welfare, and their role as audience.



\end{abstract}



\maketitle
\section{Introduction}
From cleaning our homes~\cite{Schneiders:Domestic:2021}, to mowing our lawns~\cite{Verne:Garden:2020}, to delivering shopping~\cite{Valdez:2021:Starship} and couriering items around hospitals,~\cite{Tornbjerg:Hospital:2021}, robots are finding their place in daily life. As they do so, they will inevitably interact with and be encountered by animals. These might be companion animals, the pets who share our homes or the guide dogs who help us navigate public places, but they might also be wildlife. Often these encounters will be unplanned and secondary to the robot's intended task, for example cats riding Roombas\footnote{https://youtu.be/uGI8Od22WM4}, guide dogs being confused by delivery robots~\cite{Bhat:2022:GuideDog}, or hedgehogs having to navigate in a world inhabited by lawn mowing robots~\cite{rasmussen2021wildlife}. However, they could also be intentional. We could design robots to serve animals too. Despite the inevitability of such encounters, planned or otherwise, little is known about how to design robots for animals. Can we even trust them with each other?

We present Cat Royale~\cite{BlastTheory:CatRoyale,Schneiders:2024:Interactions,Schneiders:CatRoyale_TAS:2023,Benford:2024:CatRoyaleEthics,Schneiders:2024:altHRI}, a creative exploration of designing a domestic robot to enrich the lives of cats through play. A robot, in contrast to non-embodied technology, offers the advantage of physical manipulation of objects in the enclosure. Therefore, it lends itself well to the context of cat play as it is able to, e.g., wave feathers, provide treats, or drag strings around the enclosure for the enrichment of the cats. Cat Royale was an artist-led project ~\cite{Benford:ArtistLed:2013} that set out to tackle the wider question of trust in autonomous systems. The idea of creating a playful robot for cats emerged as a means to achieve this goal, while also providing benefits for the cats. Over the course of eighteen months, the artists, in partnership with a team of HCI/HRI researchers, and supported by experts in cat play and welfare, and animal-computer interaction created a bespoke enclosure, a `cat utopia' in their terms.  At its centre, a robot arm manipulated a variety of toys to try and entertain a small community of cats who lived there for six hours a day for twelve days.  During this time, the robot offered over 500 play activities for the cats. In the background, the artists, roboticist, animal-computer-interaction specialists, machine learning experts, and animal behaviour experts worked tirelessly to ensure their safety, wellbeing and hopefully, pleasure.

While Cat Royale was undoubtedly a unique experience, and not one that can be directly transferred into people's homes, and while an art project is an unusual approach to designing robots and animal-computer interactions, we propose that Cat Royale offers important insights into the design of robots for companion animals. The artists' meticulous attention to detail in delivering a safe and entertaining experience for the cats drove them to `go beyond' many of the conventional aspects of robot design. Yes, they had to choose and adapt robot hardware and implement software for motion control, tracking the cats, and recommending games. However, they also had to carefully embed these within a wider environment that allowed the cats to engage appropriately, while humans could oversee safety and enjoy the spectacle. In short, they needed to create an entire `robot world'.

We reflect on the design of Cat Royale, and how the experience unfolded with the cats, to draw out wider implications for creating robot worlds (or, more formally ecologies~\cite{Lyle:2020:Ecology}). We distil two key contributions. First, we highlight the importance of world design for robots; of designing key aspects of the places they inhabit alongside companion animals and humans, including providing safe spaces and observation spaces for all parties, enabling robots to flexibly interact with passive objects, and finding a decoration scheme appropriate to all. Second, we reveal how designing a robot to autonomously engage animals requires consideration of interspecies interactions, from spectating, to overseeing, to cleaning up the resulting mess. Our aim is to sensitise designers to the complexities of robot encounters with animals---intended or otherwise.

\section{Related Work}
Related work falls into two categories. First, with the immediate goal of Cat Royale in mind, we review previous attempts to design digital technologies to play with cats. Second, we consider a body of work that speaks to the wider idea of designing robot worlds and digital ecologies that has emerged from our reflections.

\subsection{Playful digital technologies for cats}
\subsubsection{Commercial cat toys}
Cats are specialised meat-eaters, who have evolved to hunt and consume small prey, for which their sensory, cognitive and physical characteristics, and behavioural aptitudes are supremely adapted, including for example a high visual~\cite{miller2001vision}, auditory~\cite{Populin:1998} and tactile~\cite{bradshaw2012behaviour} sensitivity to movement. Many commercial products for cat care and entertainment are designed to leverage these characteristics and aptitudes to enhance cats' domestic experience as well as humans' experience with their feline companions. There is a significant global market for cat toys and games (valued at around US\$ 1 billion~\cite{FMI}) spanning a great variety of mechanical devices: from simple canes to which feathered objects are stringed for cats to paw at and catch as the cane oscillates under their blows (e.g.~\cite{PurrPerfect}); to complex modular circuits featuring perforated tunnels through which balls run for cats to ambush and catch at the openings located along the tunnel tracks~\cite{Catit}. A wide range of smart toys can also be found on the market: from smart boxes that activate when cats approach and that randomly poke feathers through holes for cats to try and grasp before they disappear again (e.g.~\cite{FunBox}); to wheeled mice that, after periods of inactivity, wake up and zoom away at speed, avoiding obstacles and resetting themselves as they roll over, so they can continue to entice their feline users to chase them (e.g.~\cite{Pixie}).

\subsubsection{Digital screens for cats (and humans)}
Despite the size and technological advances of the cat toy and games market, Animal-Computer Interaction (ACI) research on computing-enabled toys for multispecies play with cats has been relatively limited. One notable exception concerns exploring how digital screens may enable cats and humans to share play experiences. (e.g.,~\cite{Noz:2011:CatCat,Kasuga:2017:AnimalRobotHuman,Mancini:CHIEA:2013}). \citet{Noz:2011:CatCat} developed and evaluated a game application for iPad, designed to leverage cats' chase drive, and to enable humans and cats to play together. Its interface featured a mouse icon which ran randomly across a screen, producing sounds and changing orientation, velocity and direction when tapped; an associated application for iPhone also allowed the human to control the virtual mouse's motion variables. Human participants in a user evaluation reported that the game had been enjoyed by the cats and that it had helped them bond with them. However, these findings were based on the human participants' perceptions of the cats' experience, rather than on expert behavioural analysis over an extended period. A similar tablet game was developed by~\citet{Westerlaken:2014:Zoomorphism} to enable humans and cats to play together via an interface presenting a virtual aquarium where fishes swam around randomly for the cats to catch. The game provided both visual and audio feedback for the cats and the interface allowed the human player to modify some of the game's variables. Consistent with cats' sensory characteristics and behavioural aptitudes~\cite{Haddon:2023:Cat}, behaviour analysis of cats' interaction with the artefact suggested that the cats were particularly stimulated by the audio feedback and by variations in the speed of the fishes, and that they especially focused on the sides of the tablet when the fishes disappeared from that edge of the screen. However, while this kind of virtual experience raised the interest of many cats, researchers have warned that digital interactions which stimulate biologically important behaviours (e.g., hunting) and promise biologically significant outcomes (e.g., catching prey), without providing the release that a physical interaction would afford, may have a negative impact on the animals' welfare~\cite{Haddon:2023:Cat}.

\subsubsection{Intelligent environments and robots}
Previous work by~\citet{Pons:2015:Playful} aimed to develop intelligent, multi-modal playful environments which leverage machine learning and embodied interactions, involving multiple technological elements, to engage cats through the provision of adaptive experiences based on their responses. To this end, the researchers developed a depth-based tracking system, using sensor data from Microsoft Kinect, for detecting the location, body posture, gestures and field of view of feline and other players within the environment~\cite{Pons:2015:Playful}. They also conducted an observational study of cats interacting with different technological objects, both digital (ground projections of mice) and physical (robots Sphero and Jumping Sumo)~\cite{Pons:2017:Animals}. The study aimed to assess the system's ability to recognise cats' interactions with the stimuli, which it did with varying degrees of success. The authors also assessed the cats' responses when presented with the multimodal stimuli and found that the cats were more interested in tangible ones than they were in digital ones. 

We can consider general purpose robots to be a form of intelligent environment, especially ones that learn and adapt. In the case of Animal-Robot Interactions~\cite{Kim:ARI:2009}, whether or not animals are the target users, pursuing this aim would require robots to correctly interpret animals' communication modes, allowing them to provide input, and to respond safely and appropriately according to the needs and wants animals might express. But, in order to provide enriching experiences that can foster cats' wellbeing, the implementation of such intelligent environments might require the collection and triangulation of large amounts of data from multiple sources, informed by appropriate scientific knowledge on cat behaviour and welfare. In this regard,~\citet{Lawson:2015:Upstream} have highlighted how an imprudent use of quantifying technologies may lead to an incorrect interpretation of and response to cats' welfare needs, as well as to a weakening of the human-cat bond. The authors conducted a qualitative study with cat owners and cat welfare experts who had been asked to consider speculative designs of cat-quantifying technologies; they found that cat owners were desirous and willing to trust information on their animals provided by the technologies in question, even though this had little scientific validity. This raises the question as to what might induce people to trust these technologies and, more importantly, what conditions these should meet before they can be trusted to have a positive impact on animal guardianship, care and wellbeing, which is a key aim of the Animal-Computer Interaction (ACI) field~\cite{Mancini:ACI:2011}.

Cat Royale occupies a distinct position at the intersection of this body of related work. It responds to the importance of tangible play and the market for interactive physical toys, but aims to extend their possibilities by having an autonomous robot deploy the toys, albeit with a human-in-the-loop~\cite{Kim:ARI:2009} to moderate the robot's decisions and ensure safety. It also embeds this robot, toys and cats into an intelligent environment that monitors the resulting play and tries to learn the cat's individual preferences. Finally, it recognises questions of trust, exploring these through dialogue with its public audience.

\subsection{Robot worlds and digital ecologies}\label{sec:RW_RobotWorlds}
A key reflection that emerges from our experience of Cat Royale concerns the importance of carefully designing the wider world within which a robot must operate beyond designing the robot itself.

There is much research in Human-Robot Interaction (HRI) on investigating and designing direct dyadic interactions between humans and robots (e.g.,~\cite{AndersonBashan:2018:Greeting,Schneiders:2022:THRI}). While this will remain important, here we focus our attention on the far fewer examples of also designing the world within which such interactions occur. The Moxi robot by Diligent Robotics~\cite{moxi} is an assistive technology deployed in a variety of hospitals, all with different spatial constraints. A key challenge for Moxi concerns opening the various doors it encounters. The solution was to customise the environment, deploying robot-readable~\cite{Higginbotham:2019:RobotWorld} stickers at doors to allow Moxi to interact correctly. Another example is the SN robot, designed to mediate social interactions between lonely individuals, and ultimately reducing the feeling of loneliness~\cite{Jeong:Fribo:2018}. A robot who, through sensors placed within the environment, detects activities performed in the household (e.g., opening of doors or using the washing machine). More generally, \citet{Jeong:Fribo:2018} show how the design of the environment, and not `just' the robot design, impacts robots' capabilities, ultimately informing the entire interaction. In a similar vein, \citet{Higginbotham:2019:RobotWorld} writes of the challenges of robots and other emerging devices changing faster than the environments in which they are deployed and the infrastructure on which they depend and argues for the importance of redesigning both to create a 'robot readable world' (using a phrase credited to designer Carla Diana)~\cite{Higginbotham:2019:RobotWorld}. 

Looking beyond robots, various researchers have drawn attention to the need to understand and design interaction as part of a wider context within which it is situated. The concept of an `ecology', in various forms, has often been raised as a way to describe this broader perspective beyond the immediate device (e.g.,~\cite{Raptis:2014:Ecology,Bodker:2016:Ecology}).
In reviewing 129 works on ecological approaches to digital technology,~\citet{Lyle:2020:Ecology} identify four key types: information ecologies as systems of people, technologies, practices and values in a local environment; device ecologies as systems of interconnected devices; artifact ecologies as foregrounding the relationship of artefacts to practices; and communication ecologies as foregrounding on communication practices. They further propose that ecologies can be considered at three levels of scale: the micro scale of individuals, artefacts and tasks); the meso scale of people and practices; and the macro scale of organisations and activity systems. Others have employed the term `ecosystem' to describe the wider organisational contexts within which digital technologies exist, including the regulatory and governance structures that underpin trust~\cite{stahl:2022:computer}.
In short, consideration of the wider ecology within which technology exists can take numerous shapes and forms, but in its essence involves looking far beyond the design of `just' the technology.

\begin{figure*}[h]
    \centering
    \includegraphics[width=\linewidth]{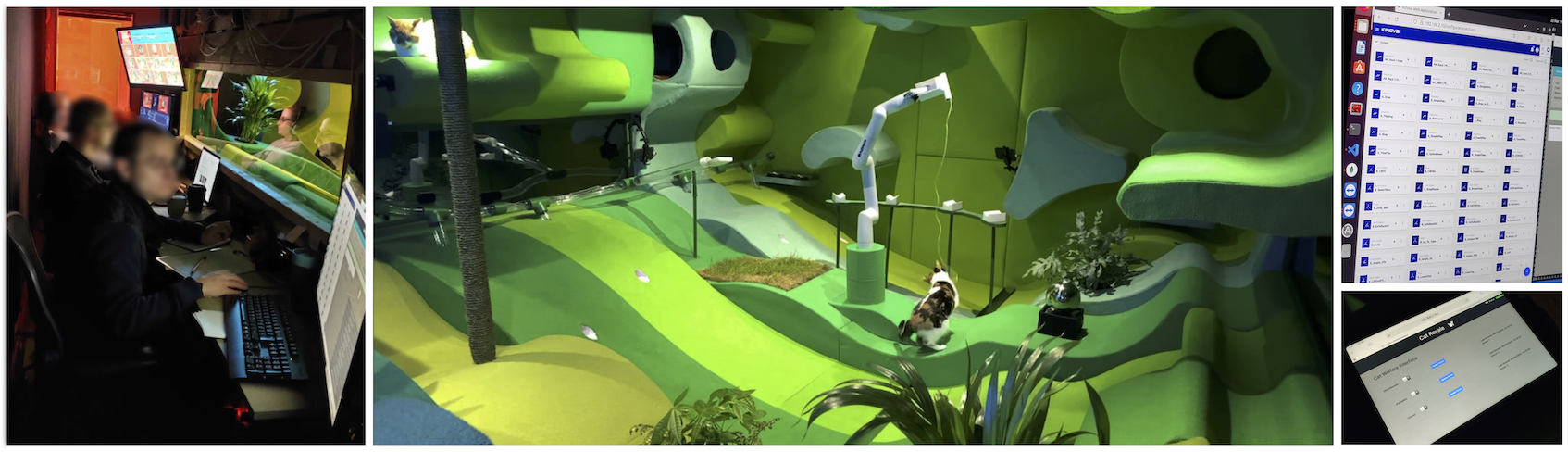}
    \caption{
    \textit{Left:} The control room (front to back): robot operator, artists, vision mixer. 
    \textit{Centre:} Inside the Cat Royale enclosure, in the centre of the room is Clover playing with the \texttt{String} deployed by the Robot. In the upper left corner Pumpkin is resting on one of the high perches. Behind the robot are the four toy racks used to store end-effectors. Two of the high dens are visible, allowing the cats to retreat. On the left, a scratching post, as well as the ball run tube system is visible. Within the enclosure a multitude of plants are distributed.
    \textit{Right top:} The robot control system. The activity library is on the left. Each blue box (52 visible) represents one of the many possible robot actions. 
    \textit{Right bottom:} The iPad based interface for documenting the cat stress score~\cite{kessler:turner:1997}, this is based on a seven point scale (1: Fully Relaxed to 7: Terrified). 
    }
    \label{fig:CRSystem}
\end{figure*}

Turning to the home as a distinctive ecology, the architect Stewart Brand offered a framework to express the evolution of a building over many years in terms of so-called `shearing layers'. He presents six distinct layers: 1) the physical `site' characterised by the geographical site which defines the boundaries of the structure; 2) the `structure', which is the building or structure itself; 3) the 'skin', which is embodied in the facade and walls of the structure; 4) the `services' that are embedded into these, including wiring and other infrastructure needed to operate and maintain it; 5) the `space plan' which is the layout of the space including placement of walls, doors, and floors; as well as 6) and the ephemeral 'stuff' such as furniture and decorative items which is less static and permanent~\cite{brand:1995:buildings}. \citet{rodden:2003:evolution} build on this framework to consider how digital technologies become integrated into and co-evolve with the home. While digital technologies might impact on (and be impacted by), all six layers, skin, services, space plan and stuff are most immediately obvious to consider in the design and deployment of any new technology. Moreover, they may come into conflict with each other due to the different rates at which they evolve, consistent with Higgenbotham's line of argument highlighting the tension between robots, robot environments and infrastructure~\cite{Higginbotham:2019:RobotWorld}.

\section{Approach}
With Cat Royale, we have followed the approach of Performance-led Research in the Wild~\cite{Benford:ArtistLed:2013} as a distinctive way to engage HCI research with the performing arts to the benefit of both. Falling under the broad umbrella of Research Through Design, this is design-led, unfolding through the practical creation of public performances and installations and subsequent reflection on these to generalise design knowledge. It also sits within a broader history of HCI engaging the arts, both to address their distinctive challenges with regard to engagement, presence and other matters~\cite{edmonds2014human}, but also to surface fresh perspectives for the field, such as: provoking interpretation through ambiguity~\cite{Sengers:2006:DesignEval}; deliberately designing uncomfortable interactions~\cite{Benford:2012:UncomInter}; promoting environmental sustainability through activist art~\cite{DiSalve:2009:SusHCI} and various provocations arising from artistic speculative design~\cite{Andersen:2018:DisruptiveImprov}.

Papers grounded in Performance-led Research in the Wild have appeared in HCI for more than 20 years with the approach being explicitly named and articulated in 2013~\cite{Benford:ArtistLed:2013}. Its distinctiveness arises from the particular relationship it establishes between art practice and HCI research. Typically, the process begins with a team of professional artists and computer scientists, including HCI researchers, building a relationship. At some point the artists propose a creative vision and/or opportunity to create a new artwork. While this may be in part grounded in the technical capabilities of the researchers, and perhaps even broadly inspired by their research (especially if the collaboration is partly funded by a wider research grant), the vision is essentially an artistic one, and often takes the technology in an unexpected direction, sometimes offering a critical perspective on it. The first role of the researchers is to help the artists realise their vision, collaborating with them to iteratively develop software and hardware, and enable early public performances. In return, the researchers get to study the artists’ rationale and process, and the audience's experience. Reflection across these often yields unusual perspectives on interaction.

\begin{figure*}[t]
    \centering
    \includegraphics[height=7.4cm]{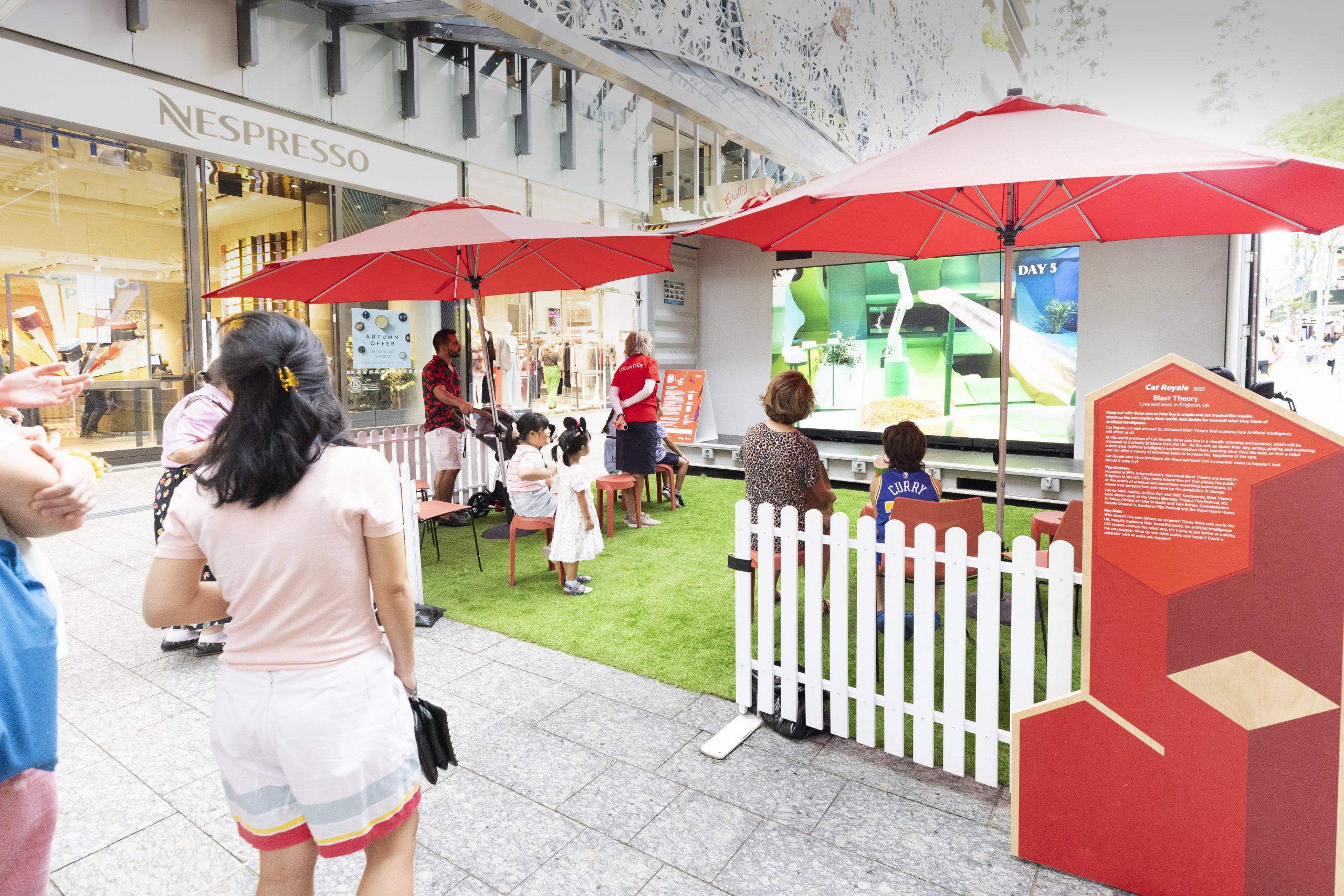}
    \hspace{.001\linewidth}
    \includegraphics[height=7.4cm]{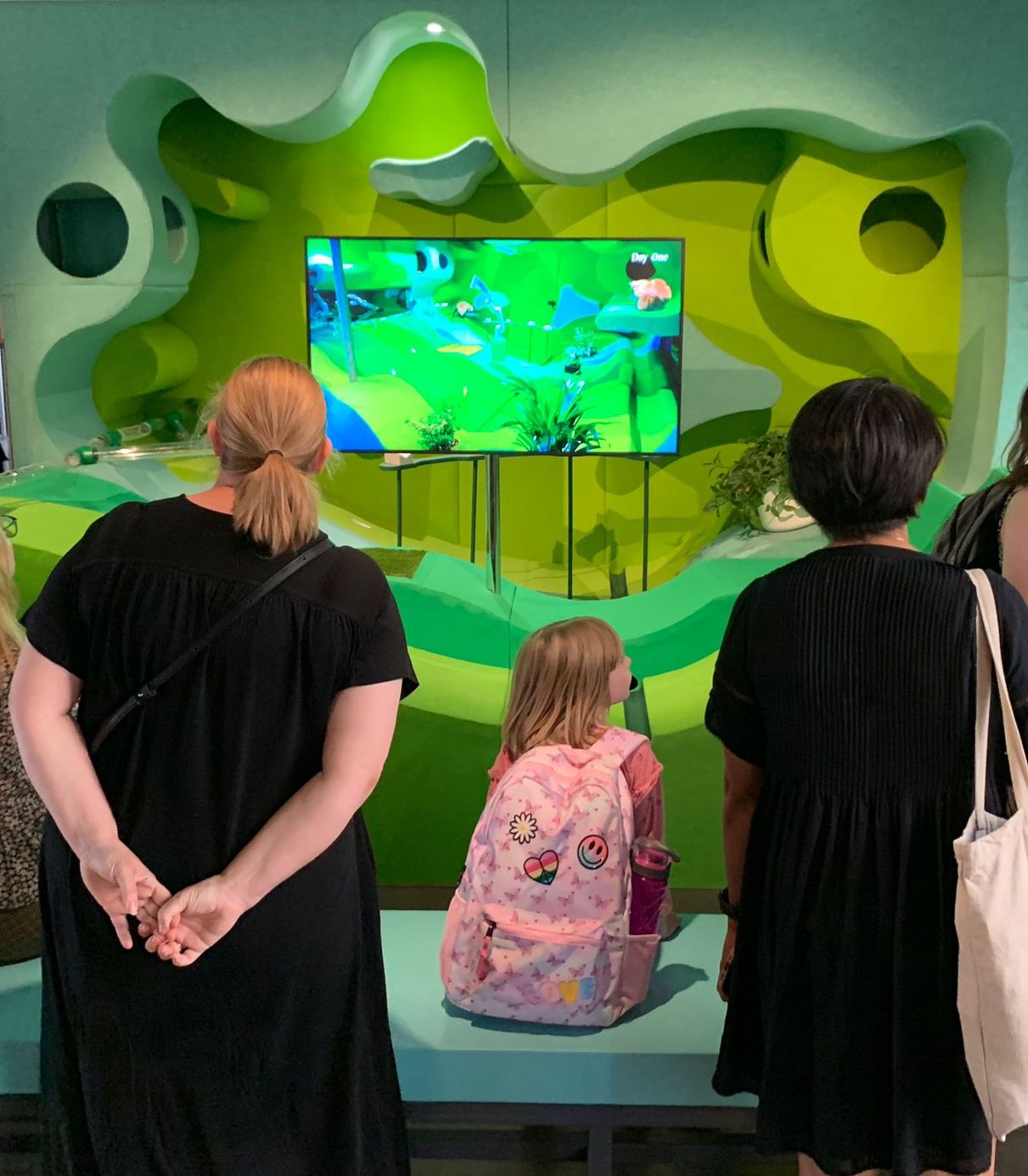}
    \caption{\textit{Left:} Public installation as presented during Curiocity Brisbane World Science Festival$^5$ (day five). \textit{Right:} Installation currently on display at the National Science Gallery in London (June 2023 - January 2024).}
    \label{fig:PublicEvent}
\end{figure*}

In this case, the artists---who have been collaborating with some of the authors on multiple projects for decades---were invited to make some form of artwork that would explore the question of trust in autonomous systems. The artists have over 30 years of experience using new technologies to create a variety of artworks, including performances, games, films, apps, and artistic installations. Following exploratory workshops at which they experienced various robots, their artistic response was Cat Royale, which the researchers then helped implement over a period of 18 months. The following account draws on design documents, meeting notes, observations, and autoethnographic field notes and video recordings (especially from the `robot operator') to reflect on the process, up to and including what happened when the cats played with the robot. Observations and autoethnographic field notes used to give the account of Cat Royale were made by various members of the team including the artists, the robot operator, and members of the research team who observed the activities taking place. This material does not cover public reaction to watching the final edited film, which is still being deployed in museums and galleries and will be reported in subsequent papers. Instead, the contribution of this paper is to report on the lessons that emerged through designing a unique and challenging artefact.

Our project was advised by, and received ethical approval from, three ethical review boards at the lead researchers' University. The Animal Welfare and Ethical Review Body (AWERB) ensured compliance with the regulations and law governing research involving animals in our country; the Committee for Animals and Research Ethics (CARE) in the School of Veterinary Science shaped and approved our cat welfare protocol. The Computer Science research ethics Committee (CSREC) approved the human-facing aspects of the project including consent, anonymity and data collection and management~\cite{Benford:2024:CatRoyaleEthics}.
\section{Cat Royale: The Multispecies World}\label{sec:CR}
To investigate multispecies interaction between human and non-human actors with a (semi)autonomous system, the artist-led Cat Royale project team, comprising artists and a multidisciplinary set of researchers, asked itself the following question:
\begin{quote}
    \textit{``How do we situate an autonomous system within a wider environment, which is simultaneously engaging for spectators, ensures cat well-being, and is suitable for the autonomous system to operate in?''}
\end{quote}

This section outlines the project and its non-human and human components and actors. The design process and the accompanying challenges of Cat Royale, which this paper focuses on, were developed over the course of eighteen months.\\


\noindent
\textit{The Enclosure.} The project's centrepiece was a bespoke enclosure, built for and inhabited by three cats. The three cats, a parent and two offspring, spent six hours a day (two periods of three hours) over twelve consecutive days within the enclosure, a total of 72 hours. As we describe below, the enclosure was designed, in the artists' words, to be a 'utopia' for the cats, a luxurious space that could cater to their various needs, including play. Its striking visual design was also intended to convey the idea of luxury to the audience when filmed via the eight cameras embedded inside.

\noindent
\textit{The Robot.} Placed in the centre of the enclosure was the robotic arm, whose sole purpose was to play with Ghostbuster, Clover, and Pumpkin, the three inhabitants of the enclosure. Provided with access to a series of custom end-effectors and a wide range of different toys and attachments, the robot was able to offer a variety of games and treats. At regular intervals, the robot's underlying system proposed a robot action aimed at maximising a specific cat's assessed happiness (see~\Cref{sec:Humans}). If approved by the lead artist of the installation, the robot would execute a playful activity targeted at increasing the cat's happiness, while being monitored by the robot operator to ensure safe robot movement in close proximity to the cats. To further ensure animal welfare, each cat's individual stress score~\cite{kessler:turner:1997} was ranked at 15-minute intervals by the team's specialist in feline behaviour. The robot performed more than 500 activities over the course of the 12 days, ranging from providing small cat treats\footnote{Regular food and water was available to the cats at all times, regardless of their engagement with the robot.}, to simple toys like a cardbox or a string, all the way to more elaborate toys such as an orange bird toy or a battery powered wiggling fish toy. Following each activity, each cat's engagement---as a proxy for positive attitude towards the robot's provided entertainment---was ranked using the Participation in Play scale~\cite{Ellis:2022:Pip}. The task executed (i.e., `present a prey game with a toy bird'), the targeted cat, and the impact on happiness where subsequently fed back into the robot's machine learning system (i.e., the decision engine) informing future proposed activities.\\

\begin{figure*}[t]
    \centering
    \includegraphics[width=\linewidth]{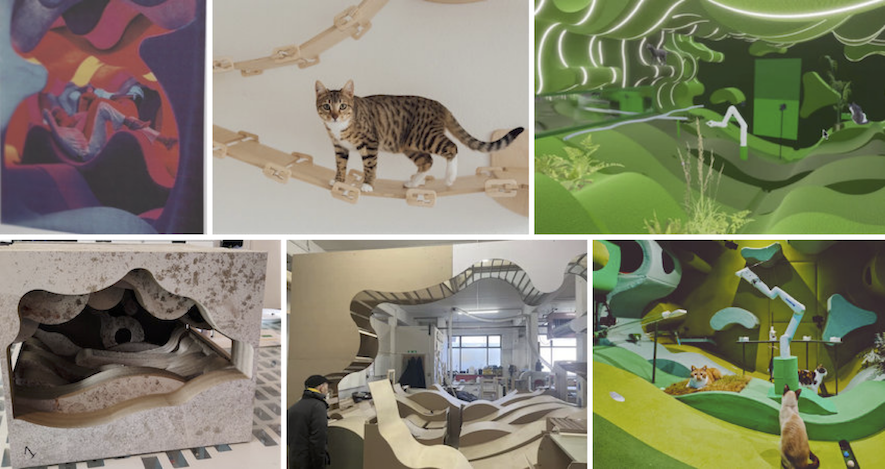}
    \caption{\textit{Top left to bottom right:} A selection of iterations showcasing the enclosure design from initial inspiration by the art of Verner Panton, to 3d renderings and scale models, to the final full sized installation.}
    \label{fig:EnvDesign}
\end{figure*}

\noindent
\textit{The Installation.} All cat and robot activities within the enclosure were constantly filmed using eight cameras (see~\Cref{fig:System_Closeups} 1), the output of which were mixed in real-time by an experienced television vision mixer to capture the best possible view of the action and to add descriptive captions to it. The footage was subsequently edited into an eight-hour long film that was exhibited to public audiences at the Curiocity Brisbane World Science Festival\footnote{https://www.worldsciencefestival.com.au/news/cat-royale-twelve-days-three-cats-one-ai-trained-robot} (see~\Cref{fig:PublicEvent} left), and at the National Science Gallery in London\footnote{https://london.sciencegallery.com/ai-artworks/cat-royale} (see~\Cref{fig:PublicEvent} right) for six months. Short highlight videos (2 -- 5 minutes) were also edited and released online during filming.

The remainder of this section, elaborates on core elements of the design of the cat and robot world that is Cat Royale. 

\subsection{Designing the enclosure}
Inspired by art from the 70s, e.g., Verner Panton Visiona II exhibit\footnote{https://www.verner-panton.com/en/collection/visiona-2/}, Cat Royale's vision was to create a luxurious environment using organic shapes and colours, which would be of interest to and appropriate for the cats that would use the space, as well as of interest to the audience who would watch the cats in the space. The design of the enclosure was informed by cat-centric criteria to provide for their needs, entailing the presence of high perches, resting dens and viewing-platforms, suspended walkways, rolling floors (which provided hiding places to ambush the toys during prey-behaviour games with the robot), and soft---and claw-able---textures. The audience's input, provided through the audience advisory panel (see~\Cref{sec:Humans}) and the artists' vision for the enclosure , had to be aligned with the cats' needs, consistent with advise from the animal welfare and behaviour experts (i.e., a veterinarian as well as an expert in feline behaviour).

To this end, the enclosure went through numerous iterations, sketches, mock-ups, 3D renderings, physical scale models, until the final design was installed (see~\Cref{fig:EnvDesign} for a selection of different stages of the enclosure design). With the enclosure being a crucial part of the success of the installation, we had to make sure that it allowed the cats enough agency to prioritise their well-being by, e.g., allowing them to engage with the robot or withdraw from it, using the many perches, walkways, dens and other resources provided, consistent with ACI research ethics guidelines~\cite{Mancini:2017:AnimalEthics}. Furthermore, apart from being entertained by the robot, the cats had species-specific and individual needs, identified by experts in feline behaviour, veterinarians and their owner. Accordingly, the enclosure was equipped with numerous cat friendly plants, a scratching post, a ball run, and a water fountain as forms of enrichment, as well as multiple feeding stations and secluded litter boxes, to prevent resource-based competition among the cats. While the enclosure was built for the cats as primary stakeholders---making traversal on the sloped surfaces suboptimal for humans---accessibility for humans was still vital. Following each three-hour session, members of the team had to be able to enter the enclosure, after the cat owner had safely removed the cats, in order to perform regular maintenance, such as vacuuming, refilling the water fountain, emptying the litter boxes, or making smaller adjustments to the robot, cameras, or enclosure. In addition to the entrance to the enclosure, two-way mirrors on both sides allowed for viewing from both the control room and the viewing area.


\begin{figure*}[h]
    \centering
    \includegraphics[width=\linewidth]{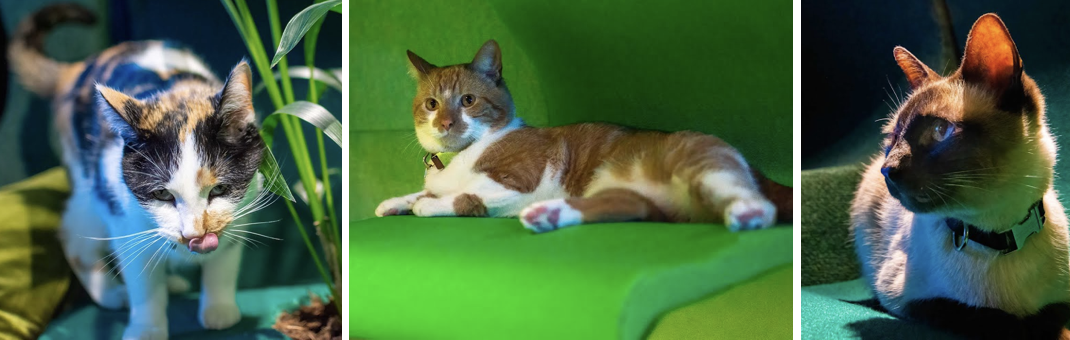}
    \caption{From left to right: Clover, Pumpkin, and Ghostbuster.}
    \label{fig:Cats}
\end{figure*}

\subsection{The Cats}
As non-human stakeholders, such as cats, are increasingly encountering autonomous systems, and vice-versa, it is important to understand how to design these systems for, and with, them. The cats' involvement was an ongoing process, from the recruitment of the specific animals, i.e., Clover, Pumpkin, and Ghostbuster (see~\Cref{fig:Cats}, left to right), to their safe return home following the conclusion of the installation, all the way making sure that they thrived.

Animal welfare was a constant priority throughout. To ensure this, among other considerations, the cats needed to be familiar with each other (so they would be comfortable sharing the enclosure) and live within close proximity of the art studio in which the enclosure was to be built, ensuring that their transport would have minimal impact.
Furthermore, in order to make sure that the cats were not deprived of contact with their owner, the owner needed to be able to move to the art studio for 17 days, including the twelve days of the installation's deployment and a five-day habituation period prior to the deployment.
Moreover, the capacities of the underlying computer Computer Vision (CV) System needed to be considered. To this end, three visually distinct cats were chosen to improve the ease of detection for the CV System. In line with ACI ethics principles proposed by Mancini~\cite{Mancini:2017:AnimalEthics}, we sought two complementary forms of consent for the feline participants. Firstly, we secured \textit{mediated} consent from the cats' owner, as the agent with the authority to represent the best interests of the cats, and secondly by designing the enclosure and the robot (as well as its placement) in a way that allowed the cats to provide or withdraw their \textit{contingent} consent, by choosing to either engage with the robot, and how, or not to engage with it and instead retreat to the many protected and comfortable spaces provided by the enclosure.

\subsection{The Robot and its Underlying Systems}\label{sec:TheRobot}
Just as the cats' recruitment was vital, so was the selection of the robot used in the enclosure. The robot's morphology, range, movement speed, and lifting capacity endowed, or deprived, the artist of different affordances in relation to the movements that could be executed. Furthermore, the robot, as instantiation of a physically embodied autonomous system, had a potentially tangible impact on the cats' wellbeing and safety. For Cat Royale, the choice was to use a lightweight robot with a small payload and reach, thereby allowing the cats to easily withdraw from it. The choice for the project fell on the Kinova Gen3 lite robot arm\footnote{https://www.kinovarobotics.com/product/gen3-lite-robots}, a small collaborative robot with 6 degrees of freedom, 0.76 meters reach, an integrated two-finger gripper providing it with a maximum payload of only 0.5 kilograms. 

Alongside the hardware, three software components were developed to drive the robot: 1) the above-mentioned CV System, 2) a decision engine, and 3) a robot control system.

\begin{figure*}
    \centering
    \includegraphics[width=\linewidth]{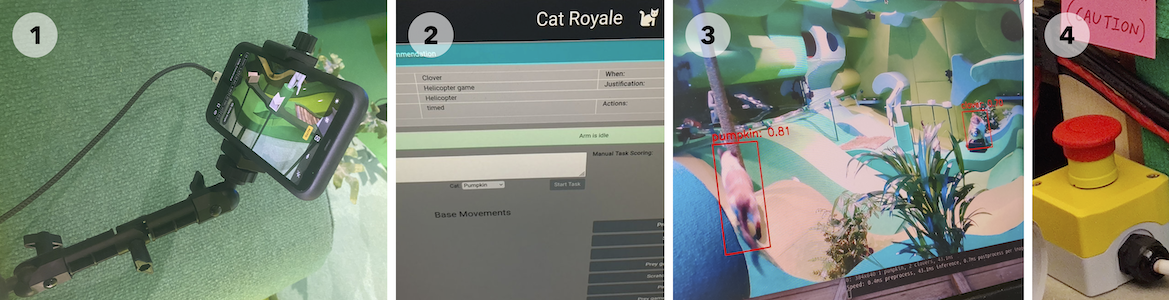}
    \caption{
    \textit{1.} One of the eight iPhones mounted in the enclosure. 
    \textit{2.} Close up of the decision engines recommendation. Visible is the proposed cat (Clover), the task (Helicopter game), the toy (Helicopter), as well as some meta data for the log files. 
    \textit{3.} Computer Vision System identifying Pumpkin and Clover, each one highlighted with a boundary box. 
    \textit{4.} Emergency stop button desk mounted next to operator.}
    \label{fig:System_Closeups}
\end{figure*}

The CV System identified the individual cats (see~\Cref{fig:System_Closeups} 3) and their positions in the enclosure so that the system could target games at specific cats. The data set that was used to train the system for this classification task included 7353 videos of cats, labelled by volunteers using the Zooniverse crowd-sourcing platform\footnote{https://www.zooniverse.org/projects/blasttheory/cat-royale}. The \textit{decision engine} was the core system which provided the robotic system with autonomy. At regular intervals, approximately six minutes after the previous activity was completed, the decision engine would propose an activity for the robot targeting any of the three cats, e.g., `\texttt{Prey game with bird colours, target: Ghostbuster}'. To ensure that a balance was struck between offering the preferred activities and providing sufficient activity variation, the decision engine 
varied the so-called exploration-to-exploitation ratio throughout the course of the project. That is, during the early days, when no or little training data existed to determine the performance of the robot's activity, the system would rely on exploration, i.e., trying random activities to collect training data. With the increase in collected training data, the system learned which activities had a positive impact on each of the cats' happiness score, allowing the decision engine to propose specific activities and toys depending on the individual cats' preferences. This led to a higher degree of exploitation, i.e., the use of specific activities that were proven to have a positive impact on cat happiness. The robot, once the specific activity was approved by the artists, would execute the movements. This was done by the robot operator (see~\Cref{sec:Humans}) using the \textit{robot control system}. The control system acted as the interface between control room and robot, and allowed the robot to autonomously plan trajectories for moves, which could be executed---under human supervision as discussed below---at will.




\subsection{Toys, Racks, and Robot Movement}
\begin{figure*}[h]
    \centering
    \includegraphics[width=\linewidth]{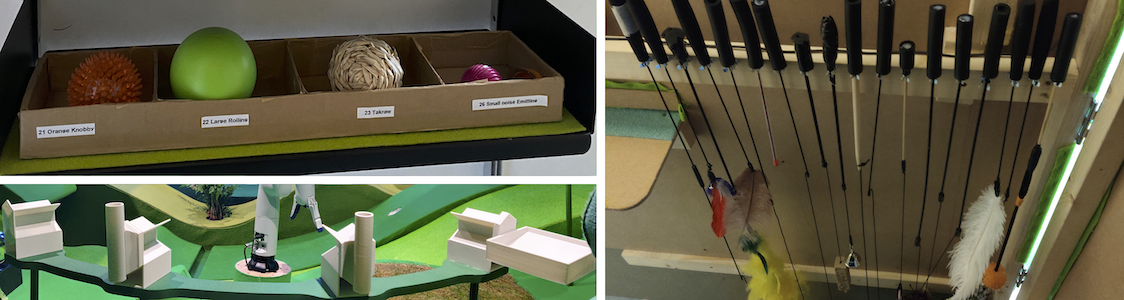}
    \caption{\textit{Left top:} Four of the wide variety of balls for available to the robot. \textit{Left bottom:} The toy rack with four end-effectors used for balls and treats (rack on the right) as well as attaching various toys to (three racks on the left). \textit{Right:} Various feather toys which can be used with the three most end-effectors.}
    \label{fig:RackAndToys}
\end{figure*}

With the robot selected, the next step was to provide it with additional affordances so that it could play a variety of games. These needed to allow it to manipulate a wide range of items (e.g., feathers, pillows, treats, bells, or balls) in order to attempt to engage the cats (see~\Cref{fig:RackAndToys}). To provide the artists with greater artistic freedom, we designed and 3D printed custom end-effectors. These were supports to which a range of objects could be safely secured and which the robot arm's two-finger gripper could securely grab, from custom-built racks, and hold. This way, the end-effectors augmented the robot's capabilities beyond the limitations of the two-finger gripper, enabling the artists to prototype and experiment with different toys and movements, expanding the repertoire of playful experiences available to the cats---including some that allowed the cats to, in some regard, overpower the robot. Thus, the customisation allowed a wider range of creative possibilities for the artists and enrichment activities for the cats.
Perhaps, the most vital capacity of the robot was its movement. We needed to be able to generate movements with the robot that struck a balance between predictable routine and exciting novelty for the cats. In addition to engaging the cats, the art installation needed to captivate audiences. So, while the installation started with a set of supported robot movements, the robot operator---directed by the artists---added new movements on a daily basis, gradually increasing the repertoire of the robot. The initial set of movements included the controlled picking-up and putting-down of toys from the toy racks. During the twelve days, the movement library available to the robot grew through the addition of new ways of performing, for example, `dragging', `dangling', `dropping', `throwing', and `offering' motions. The need to add new movements on a daily basis, necessitated the robot---and its underlying systems---to allow for rapid prototyping and development of new sequences. Furthermore, it was vital to design and implement a variety of movements, as not all movements made sense with all toys, e.g., a `throwing' movement made sense with a ball, while not being applicable with a feather toy. This customisation of movements allowed the robot to engage the cats differently depending on the specific toy used, and the individual cats' preferences.

\subsection{Humans-in-the-loop}\label{sec:Humans}
While, in line with ACI research ethics recommendations~\cite{Mancini:2017:AnimalEthics}, the Cat Royale project placed the animals at the centre, a multitude of human stakeholders were involved and therefore had to be accounted for during the conceptualisation and deployment of the artistic installation. This included, the artists, two robot operators, and the audience, an animal welfare officer, a toy wrangler, the cats' owner, a vision mixer, and the members of three ethics panels who conducted the project's ethical review and approved the proposed work. While the human actors in some sense collaborated on the same tasks---i.e., ensuring the cats' welfare and increasing their happiness, and successfully delivering an artistic installation for the audience---their impact on the installation varied.

The audience had no direct impact on the execution of the project but were involved in its conceptualisation---represented by 15 members from the general public who formed an audience advisory panel. This had an impact upon the design of the environment by, e.g., conceptualising how a cat utopia would look and providing input as to how related information was to be presented to audiences around the world. The latter included, e.g., how to convey information about the cats' happiness scores and the decisions made by the robot. Through the panel's representation, the general public was given a voice in the design of the project. 

Behind the scenes of Cat Royale (see~\Cref{fig:CRSystem} left) the artists and the robot operator collaborated closely to ensure the successful operation of the robot. The system, specifically the decision engine as described in~\Cref{sec:TheRobot}, proposed concrete actions for the robot to carry out in order to increase the happiness of the cats. It was then up to the directing artist to make a judgement call---trusting and accepting the system's recommendation, executing the proposed action and hoping it would lead to increased happiness for the cats and excitement for the audience; or rejecting the proposed action and requesting a new one. 

To ensure safety, the system relied on a human-in-the-loop mediating between artist and robot, namely a robot operator. The robot operator would, following the decision of the artist, initiate and monitor the robot during task execution. Their primary responsibility was to ensure that the robot's trajectories would in no way put the cats at risk. To ensure that every action of the robot was monitored and that the operator could intervene instantly, if needed, the system relied on a dead man's switch. This was a designated button that enabled the robot to move as long as the operator pressed it but that stopped all robot movement when released. This was a safety precaution to ensure that the operator was in control at all times. 
Therefore, as the one who held ultimate control behind the scenes, the robot operator could---against the artists and the decision engines recommendation---stop any robot action, if they perceived it as posing a risk to the cats. As an additional safety measure, the robot operator had access to an emergency stop (see~\Cref{fig:System_Closeups} 4) that would cut all power to the robot.\footnote{At no point was it necessary to use the emergency stop button in order to ensure cat safety.} 

Following each performed robot activity, the system needed information regarding the cats' happiness score, i.e., the impact of the given activity on each individual cat's happiness score. To achieve this, the lead artists---who closely monitored the environment, focusing on the cats' engagement---used the Participation in Play scale~\cite{Ellis:2022:Pip} to rank each cat's engagement, which acted as a proxy for happiness on a scale from 0 to 5 (e.g., `No interest in playing or retreats from attempts at play' to `Enthusiastically playing, moving around the environment'). Furthermore, the cats' stress score~\cite{kessler:turner:1997} for each cat was recorded using instantaneous sampling (i.e., at predetermined time intervals of 15 minutes) by the animal welfare officer. At each 15-minute key point, a tablet would prompt them to evaluate the facial expression and body language of each cat (stomach, tail, eyes, pupils, ears, whiskers, and vocal behaviour), and identify the current stress score for each cat on a 7-point scale (1: Fully Relaxed to 7: Terrified).

During the entire 12-day installation period, the vision mixer, in real time, switched between the eight different camera angles and a series of overlay screens (e.g., `Pumpkin happiness score increased by 2\%' or `Coming next: Helicopter game for Clover'). This edited 6-hour daily video, cut in real time, was presented to the audience at the Curiosity Brisbane World Science Festival in Brisbane, Australia. Therefore, the vision mixer had the important function of creating the final perspectives made available to the public.
\section{The Narrative Unravelled}\label{sec:WhatHappened}
We now turn our attention to the events that transpired during the deployment of Cat Royale. While our aim in this paper is not to evaluate the success of Cat Royale in terms of effectively playing with the cats, we do note that, from the perspective of the artists and wider team, the experience was judged to be successful. The successful completion of Cat Royale, while not formally evaluated in this paper, is multilayered. Firstly, the cats safely inhabited the enclosure for twelve days without needing to be withdrawn due to stress or injury. The cats waited in front of the door to the enclosure, seemingly awaiting to gain access to it. Secondly, within the enclosure, they voluntarily chose to engage with the robot's games, 
typically descending from their perches to play for several minutes at a time. Thirdly, the team were informally (i.e., not backed up by a scientific analysis), of the view that, on the whole, the play was pleasurable and not frustrating for the cats. This sentiment was shared by the animal welfare officer based on the cats' body language, as well as the cat owner. Lastly, Cat Royale seemed enjoyable for humans to watch. With countless visitors at the installation in Brisbane, tens of thousands of views on YouTube, thousands of likes on social media, and extensive engagement in comment threads across various platforms, the project effectively captured public attention.. However, there were revealing incidents along the way. In the following, we draw on autoethnographic field notes and video recordings from the team, especially from the two robot operators and the lead artist, to paint a picture of how Cat Royale unfolded. We present a range of vignettes from typical operation to more challenging examples---requiring careful management---all of which presents the orchestration of Cat Royale.

\subsection{Typical Robot Engagement}\label{sec:success}
This section presents a cat-centric perspective by highlighting what typically happened within the enclosure. Vignette~\ref{vig:RegOccurrence}) describes a representative example of the robot activity in which some---but not necessarily all---cats engaged in playful behaviour. This type of activity, with a variety of different toys and movement sequences, occurred several hundred times over the course of the twelve days.

\begin{quote}
    \texttt{17:47:\quad} Just over halfway through the second session of the ninth day, the decision engine proposes `\texttt{Helicopter prey game, target: Ghostbuster}'.
    
    \vspace{2mm}
    \texttt{17:48:\quad} Around thirty seconds later, once the activity has been announced to the audience watching in Brisbane, the robot awakens and moves towards the helicopter toy\footnote{A small three winged propeller with feathers at the end of a string connected to a support}, which is currently mounted on the second toy rack. 
    
    \vspace{2mm}
    \texttt{17:48:\quad} With a circular motion of its first joint (i.e., the base) the robot rotates the helicopter toy towards the centre of the room. While all three cats are following the toy's movement with intensity, Pumpkin is the first to lower his posture and wiggle his entire body, indicating an impending pounce. Indeed, one minute and four seconds after the robot has picked up the toy, Pumpkin pounces on it. 
    
    \vspace{2mm}
    \texttt{17:49:\quad} After an initial period of observation, Ghostbuster joins Pumpkin in attacking the toy, all the while Clover continues to monitor the situation from the elevation of one of the resting dens located under the ceiling of the enclosure.
    
    \vspace{2mm}
    \texttt{17:49 - 17:52:\quad} While Pumpkin disengages and retreats to one of the high walkways, Ghostbuster continuously follows the toy as the robot slowly drags it around itself. As the robot lifts the toy [\texttt{17:52}], Ghostbuster jumps towards it and continuously hits it with his paws, only to disengage shortly after. Throughout the entire engagement, the cats pay attention to the toy, while being seemingly disinterested in the robot itself.

    \vspace{2mm}
    \texttt{17:54:\quad} Once Ghostbuster stops engaging with the Helicopter toy, the robot carefully places the toy back onto the rack before gracefully going back to its retracted stand-by position, awaiting the next command. Based on his interaction with the toy, Ghostbuster's happiness score is updated as having increased by 3\%.

    \vspace{2mm}
    \texttt{17:48 - 17:54:\quad} During this entire time, the various humans-in-the-loop follow the action intensely. In the control room, the vision mixer stares at the overview screen with the eight cameras, making sure that the audience is presented with the most interesting angle available. The artist observes the cats through the two way mirror, excited about every cat-robot engagement. The robot operator, whose finger is pressed onto the dead man's switch, observes the robot---and its surroundings---should they need to pause the robot's trajectory. On the other side of the installation---in the viewing area---the animal welfare officer pays attention to every little signal the cats' body language express, ready to estimate their emotional states.
    
    \captionof{vignette}{Cat Royale, a regular occurrence. Day 9, afternoon session.}
    \label{vig:RegOccurrence}
\end{quote}

The activity described above could be considered a success. Most importantly, no interventions by the robot operator or animal welfare officer were needed, as the cats' welfare was at no point at risk. Furthermore, two of the cats engaged in the activity, especially Ghostbuster, who was the decision engine's target for this specific activity. During Ghostbusters and Pumpkin's engagement with the toy, Clover continuously used the high platforms and dens to monitor the situation below. Since this activity was successful in safely engaging the cats, it provided---based on the artist's judgement---interesting footage for the video to be presented to the audience. Thus, this is a good example of an activity that achieved the primary task~\Cref{sec:CR} of ensuring cat well-being while engaging both the cats and the audience.

\subsection{Intervening with the Robots Autonomy}\label{sec:Intervention}
The movement executed by the robot happened in an autonomous (but monitored) manner. If everything went according to plan, the robot operator initiated the movement---by keeping the dead man's 
switch pressed---until the entire movement sequence was successfully completed and the toy was placed back on the rack. However, as autonomous agents, the behaviour of the cats could not be predicted nor controlled; similarly, several of the toys responded unpredictably to the cats' interaction due to, e.g., different lengths in the string to which they were attached or slight variations in the angle at which the toy had been placed onto the rack following a previous activity and prior to initiation of a new activity. Therefore, at times, there was a need for manually overriding parts of the system. This was done in several different ways, both by the artists and by the robot operator.

\begin{figure*}[t]
    \centering
    \includegraphics[width=\linewidth]{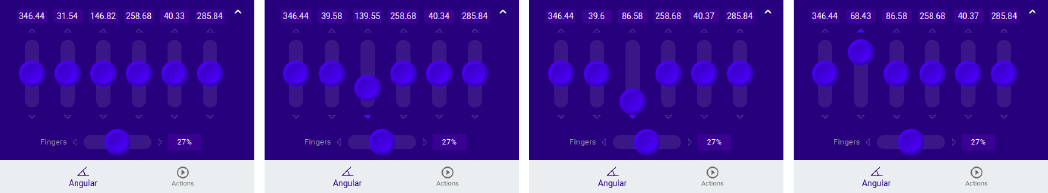}
    \caption{Robot control interface for the manual adjustments of the six degrees of freedom and the opening/closing of the gripper.}
    \label{fig:ManualIntervention}
\end{figure*}

\subsubsection{Artistic intervention}
Following the proposal of an action for a specific cat by the decision engine, the artists could chose to request a new activity, rejecting the initial proposition. This happened frequently (226 times), for a variety of reasons. 

One reason for rejecting a specific action could be that the action was being performed too frequently (e.g., `\texttt{Ping pong ball run, target: Clover. REJECTED - repetition}'). Other reasons included the lack of correct calibration of the activity, which could lead to robot failure---necessitating a restart (e.g. `\texttt{Kicking toy in animal print, target: Ghostbuster. REJECTED - `breaks' the robot arm}'). A third reason for rejection was related to the `status' of the enclosure. Dropping a ball in the ball-run, for instance, was eventually rejected as a previous ball had become stuck, i.e., `\texttt{Perforated ball tube run, target: Ghostbuster. REJECTED - ball run is blocked}'. Lastly, but most importantly, some actions were rejected to ensure animal welfare. While, the cats might have been interested in receiving frequent treats, eating too many treats would not have been beneficial for their health. Therefore tasks like `\texttt{Meaty stick treat, target: Clover. REJECTED - received enough treats this session}' were at times rejected. 
Furthermore, the artists could reject a proposed action, and instead request another specific action. This happened much less frequently (17 times) and was typically related to the desire to (i) try out a newly implemented task, or (ii) take advantage of the momentary position of one (or more) of the cats. An example of the latter occurred in the second session on day eight: `\texttt{Perforated ball run for Clover. REQUESTED - Clover is looking at the ball run.}'

\subsubsection{Operator intervention}
In addition to the artists' intervention, other types of interference with the system's autonomy could occur, namely by intervention from the animal welfare officer---which never occurred---or from the robot operator. Both could, at their own discretion, interrupt the installation or specific robot movements, if they deemed that the cats' welfare was at risk (e.g., to avoid a collision with any of the cats). Vignette~\ref{vig:manual_override} highlights one such occurrence by the robot operator.

\begin{quote}
    \texttt{12:37:}\quad Towards the end of the morning session the decision engine proposes `\texttt{Simple game with feather boa, target: Pumpkin}'. This activity has been successfully used once before the previous day, resulting in a maximum happiness score for Pumpkin.

\vspace{2mm}
    \texttt{12:38:\quad} Shortly after it successfully picks up the toy, the robot moves the feather boa in a counter clockwise fashion towards the centre of the room, rotating around its base. While passing the ball run system, which some of the smaller balls can roll through, the soft feather boa collides with one of the pipes. Continuing the movement along the current trajectory could break the feather boa or, worse, it could break the tubes which might then hit the cats.

\vspace{2mm}
    \texttt{12:38} At the onset of the collision, the robot operator releases the dead-man's switch, effectively stopping the robot's motion mid trajectory. Utilising the six virtual sliders, which are part of the control interface (see~\Cref{fig:ManualIntervention}), the robot operator is able to manipulate each robot joint in isolation, allowing them to untangle the feather boa, bring the end-effector to a safe position, and continue the trajectory of the planned task. 
    
\vspace{2mm}
    \texttt{13:10\quad} Following the morning session, the stick to which the feather boa is stringed is shortened, which prevents future collisions with the tube system.

    \captionof{vignette}{Cat Royale, manually control to recover from collisions with objects in the enclosure. Day 3, morning session.}
    \label{vig:manual_override}
\end{quote}

As Vignette~\ref{vig:manual_override} highlights, manual overrides to prevent damage to the enclosure, the robot and, most importantly, the cats were possible. However, none of these interferences---be it by the artists or the robot operator---were without consequences. Artists' rejections of specific tasks, for a variety of reasons, could lead to a shortage of training data for the decision engine on this particular activity, thereby influencing future proposed activities. In the same fashion, operator interference with toys' trajectories changed the activity. This had the potential to alter the cats' engagement with the activity and ultimately the activity's impact on their happiness score (using the PIP scale~\cite{Ellis:2022:Pip}), whether  positively or negatively. This, just as the artists' interference, had implications for the data fed back to the decision engine, thereby potentially affecting future activities. 




\subsection{The Cat is in Control}\label{sec:CatInControl}
\begin{figure*}[t]
    \centering
    \includegraphics[width=\linewidth]{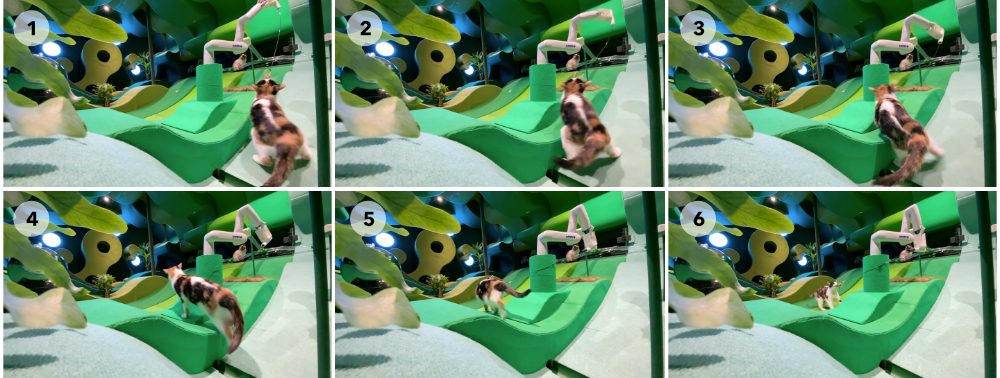}
    \caption{Day 10: Clover forcefully pulling at the right angle, leading to loosening of the joint and an error occurrence in the control interface.}
    \label{fig:Clover-vs-Kinova}
\end{figure*}
While many interactions between the cats and the robot went as expected (see~\Cref{sec:success}), at times unanticipated events occurred. One of these is the loss of human control, as it could be removed from the operator by the cats' forceful intervention. At 10:45 in the morning session on day 10, Clover discovered that, when pulling from a particular angle, she was able to physically overpower the robots joints, leaving the operator powerless for a brief moment. \Cref{fig:Clover-vs-Kinova} illustrates an occurrence of Clover's `forceful taking of control'. This encounter occurred just as the play activity was about to conclude, at least from the robots' perspective. However, for Clover, the orange bird toy was far more popular than most other toys and play activities, and she was not ready to part with it just yet. As visible from~\Cref{fig:Clover-vs-Kinova} picture 1--6, Clover pulls the Kinova while the arm's last section is positioned vertically (i.e., when it can be moved towards the floor or towards the ceiling). Following a short pull by Clover, the joint unlocks, which results in it dropping down (due to gravity). This overpowering of the robot by one of the cats has an effect well beyond the cats' enclosure: in the control room, the robot operator receives an error code that demands to be acknowledged because, during this brief period, no manipulations of the robot's joints are possible. Furthermore, when the trajectory of the robot is abruptly disrupted, the activity it was performing is cancelled and the robot has to re-initiate the proposed movement.


This---hunting or attacking the toys---behaviour persisted, with Clover ultimately being rewarded by managing to strip the robot of the orange bird toy and dragging it away. All the while, the string which connected the bird toy and the stick the robot was holding, became stuck in the water fountain, tipping it over. Unlike many digital interactions, the physical embodiment of the robot allowed the cats to `disassemble' it by taking the toys from it. This allowed the interaction to fulfil the cats' biological drive stimulated by the robot (i.e., hunting), allowing them to grab, manipulate and drag objects (i.e., prey), thus positively impacting Clovers' welfare~\cite{Haddon:2023:Cat}.

\section{Discussion - Designing Robot Worlds}

Reflecting on the artists' final design and the experience of the cats, we suggest that Cat Royale provides an existence proof that it is indeed possible to design a general purpose robot to play a variety of games that repeatedly engage cats in what appear to be naturally playful behaviours (e.g., that at first sight appear to engage their natural hunting behaviours). The cats did not appear to get bored, but rather to become confident, even assertive, in engaging the robot in such play. It may even be possible for a future robot to learn the preferences of individual cats and to adapt its play to them.

However, as we saw in the many details of the artists' design, this was far from being a simple proposition. Delivering Cat Royale required the artists to innovate solutions to a wide variety of challenges, including: choosing an appropriate robot; designing its movements; introducing systems to make these safe; developing a decision engine to recommend games; choosing suitable (passive) toys and adapting them so the robot could deploy them; building a magnetic rack storage system for the toys; creating a surrounding enclosure with spaces and facilities to make the cats comfortable and meet their other needs; appropriately decorating this for the cats, computer vision system and human audience; creating an external control room from which the play could be monitored and broadcast; and putting in place a wide variety of supporting humans roles. Furthermore, to make Cat Royale a success they had to ensure that all of these elements worked together as a coherent whole. 

What is notable about this list of activities is just how much of it extends beyond designing the robot itself or its direct interactions with the cats (though these certainly appear). Rather, the artists designed an entire `robot world'. 
This reflects the idea of understanding and designing interaction in terms of entire ecologies, not just devices, as we reviewed earlier. Cat Royale provides an extreme example of designing an entire information ecology of people, technologies, practices and values in a local environment, now extended to also include animals.
However, it is also a highly unusual example; in this art project, the artists had more or less free licence to be creative with the design of their robot world. They also had the luxury of starting with a blank page. Cat Royale seems unlikely to directly transfer to people's homes due to its size, cost, need for a humans in supporting roles, and the fact that most homes are already replete with other `stuff'. So what broader lessons can we generalise from this example. What does Cat Royale tell us about designing more everyday robot worlds for cats or other animals?

We now look under the surface of its design to distil underlying principles, grouping them under two broad themes that capture the essence of how the artists approached their robot world: designing the robot's physical world and designing multispecies interactions also involving humans.



\subsection{Designing the Robot's Physical Worlds}

In order to bring some coherent structure to the many details of designing the robot’s physical world, we return to Brand’s framework (as introduced in~\Cref{sec:RW_RobotWorlds}) for the evolution of buildings~\cite{brand:1995:buildings} as previously introduced to HCI~\cite{rodden:2003:evolution}. Cat Royale reveals how the creators of robot worlds need to pay attention to stuff, space-plan and services (as exemplified by the toys the robot wielded, the enclosure and the surrounding service space).\\

\noindent
\textbf{Designing passive \textit{stuff} to mediate interaction.} Although it may seem trivial to point out that robots will often interact with animals via passive objects, it is worth reflecting on important facets of their design. It was easy to extend the robot’s repertoire of actions by simply introducing new toys. Moreover, the movements of strings and similarly flexible materials afforded a range of natural and somewhat unpredictable movements that would have been difficult to programme. The toys were safe to lick, bite, claw and carry away. They also draw attention away from the arm itself, in the interests of both animal and robot safety. Toys simultaneously made the possibility of collisions more likely, but also much softer and safer. They could even be released into the environment to create more playful effects and again to distract the animals away from the robot. However, they introduced their own challenges: requiring a storage system that was accessible to the robot but animal-proof, while their flexibility allowed the animal to take control, making it difficult to know when and how to best release them and thus introducing a point of vulnerability---e.g., allowing Clover to `break' the robot (see~\Cref{fig:Clover-vs-Kinova}). While Cat Royale’s objects were passive, we envisage that future work could endow `smarter' end-effectors with greater agency through local sensing and actuation, for example being able to detect when they were under strain or engaging in locally autonomous behaviours~\cite{Pons:2015:Playful}.\\

\noindent
\textbf{Designing robot \textit{space plans}.} An important feature of Cat Royale's enclosure was that it provided a safe space for both animals and robot. Feeding trays, litter trays and sleeping areas were out of the robot’s reach (see Vignette~\ref{vig:RegOccurrence} and~\Cref{sec:success}). Animals could traverse the space via raised walkways that similarly avoided the robot. Importantly, they could also watch it from an elevated position, choosing when to descend and engage, which both preserved their autonomy and facilitated their spontaneous (i.e., hunting) behaviour, and drew them into engagement, in a similar manner to the Honeypot effect reported for humans using public displays~\cite{Hornecker:2007:Honeypot}. The robot too had its own safe space in the form of the raised racks where it could store the toys. A second important aspect of the space plan was the visual decoration which involved a balancing act between being aesthetically pleasing to humans (especially when seen through a camera), suitable for the cats, and also convenient for the vision system (including being able to discriminate the distinct visual markings of individual animals). Future opportunities might build on the experience of the Moxi robot and the idea of `robot readable worlds'~\cite{Higginbotham:2019:RobotWorld} by deploying visual markers that could be read by the robot while utilising aesthetically pleasing graphic designs~\cite{Getschmann:2021:Seedmarkers,Benford:2017:Markers}.\\

\noindent
\textbf{Designing robot \textit{services}.} Deploying technologies brings requirements for services such as power, cabling, networking and physical access for monitoring cleaning and repairing. Cat Royale established an extensive space surrounding the enclosure from which such services could be provided, including access hatches through which human wranglers could reach in without the animals exiting.

To generalise further, interior design---which broadly encompasses \textit{stuff}, \textit{space plan} and the visible aspects of \textit{services}---is an important part of the ecology of the home. On the one hand, it is a necessary part of making the home comfortable, maintaining it, and installing new technologies, and so must be considered when introducing a robot (and indeed an animal). As previous studies have shown, people already do this; the owners of domestic robots (e.g., lawn mowing or vacuuming cleaning robots~\cite{Schneiders:Domestic:2021,Verne:Garden:2020}) already re-structure and decorate their homes and gardens in order to make them more robot accessible. 
On the other hand, interior design can be a creative and rewarding activity enabling people to personalise their homes, and is something that many enjoy doing and are skilled at. Thus, while it might seem technically trivial to focus on interior design for robots rather than developing their hardware and software to adapt to prevailing conditions, doing so may well prove to be an empowering strategy for those who live with them. Interior design of the robot world, e.g.,~\cite{Jeong:Fribo:2018,Schneiders:Domestic:2021}, might allow for new ways of robot interactions regardless of robotic expertise.\\

\noindent
In this paper, we have presented a case in which not only we designed for interactions with the robot, but we also designed three of the six shearing layers~\cite{brand:1995:buildings}. Specifically, the \textit{stuff}, \textit{space}, and \textit{services} surrounding the deployment of the installation. We acknowledge that---in a less controlled environment---this might not always be possible; however, at the same time, we argue that considering if, and when, the shearing-layers can inform the implementation and design of digital technologies can be useful---especially in the case of physically embodied technologies such as robots. Prior examples of how to design robots and, for example, their spaces~\cite{Schneiders:Domestic:2021} to accommodate them in environments shared with human and non-human actors~\cite{Bendel:2017:ladybird} have further shown the importance of this. An extreme case of emergent interaction between animals (hedgehogs) and technology (lawn mowing robots) in a shared environment is described by~\citet{rasmussen2021wildlife}. The lawn mowing robots, designed to replace human labour, lack consideration for the non-human actors in their ecosystems leading to, in extreme cases, severe injuries to the hedgehogs with which they share an environment. To prevent negative consequences for non-human actors, we argue that it is important to consider the adoption of technology from an ecological perspective expanding beyond the human- and technology-centred viewpoints.

\subsection{Enhancing Multispecies Interactions}
Adopting an ecological perspective on computer systems foregrounds the often diverse roles played by people in these systems. The robot in Cat Royale was designed to appear, to the audience, to autonomously interact with animals without humans `being in the room'. However, we have seen (e.g.,~\Cref{sec:Intervention}) that, behind the scenes, various human roles were implicated in realising these interactions and ensuring they were as engaging and safe as possible: the robot operator, animal welfare officer, toy wrangler, artists, vision mixer, cat owner, and also the audience who watched the edited recordings. One might be tempted to argue that many of these human roles were only required due to the prototypical nature of Cat Royale, with humans initially delivering system functions that would eventually be automated. Perhaps, for example, the operator's role and the `happiness' scoring of the cats' engagement (as described in~\Cref{sec:CR}) could be subsumed by a future more autonomous AI-driven robot. Under this view, the various roles humans played in the project could be seen as revealing gaps in the technology and clarifying requirements for further development. However, our ecological approach leads us to take an alternative view: that humans, just as the three cats, are a vital part of the ecosystem and that we need to design to support and even enhance their involvement in it rather than design to remove them. The approach of designing computer ecosystems to create opportunities for rewarding engagement, rather than to reduce human labour, has previously been explored in the context of providing support at online festivals~\cite{benford2023infrastructures}; while owning animals is not formally `volunteering' of this kind, it is often `voluntary', perhaps similarly implying that we should seek to maximise opportunities for rewarding human engagement. This insight also reflects Lawson et al.'s~\cite{Lawson:2015:Upstream} concern that digital technologies developed for animals may potentially weaken human-animal bonds. With this in mind, we generalise from the specific human roles in Cat Royale to propose key directions for enriching human involvement in robot worlds.\\

\noindent
\textbf{Enriching human-animal relationships.} Cat Royale showed how both a robot operator and the lead artists needed to be closely involved in getting the robot to successfully play with cats, the former to finesse its physical movements by using the deadman’s switch to advance or stop it in its pre-planned sequence and sometimes to improvise new moves, and the latter to review its recommendations for games to play. An interesting strategy would be to extend rather than replace such roles. Might a mixed-initiative system~\cite{Allen:1999:MixedIni} that combined the strengths of a robot (precise, slow and patient movements) and human (fine judgement of animals' reactions and ability to improvise) deliver better experiences for the animal, but also for the human, involving them in creating and delivering new experiences? And might this extend to other functions beyond play? Could we similarly enhance human engagement with animal care activities such as feeding and grooming? An important topic for future research concerns better understanding how animals (and possibly humans too) might view the robot as an interlocutor in their interactions. In Cat Royale, while the behaviour of the toys clearly elicited playful responses, at this stage it is not possible to tell how the cats perceived the robot itself as an actor. However, researchers have observed the emergence of social behaviour in dogs who encountered autonomous robots~\cite{Gergely:2013:Dog}, demonstrating that at least some animals perceive robots as social actors.

Even in situations where we might want robots to interact autonomously with animals, such as when humans cannot be present, Cat Royale reveals opportunities for enhancing human offline involvement. Might the role of the audience in Cat Royale (supported by the vision mixer), be transformed into something more personal and interactive, with humans enjoying watching footage of their animals, but also categorising and rating their interactions so as to better train the robot? While we want to avoid the risks of inappropriately `quantifying' animals through over-use of tracking technologies as discussed by~\citet{Lawson:2015:Upstream}, might engaging with data to directly help train autonomous systems potentially enrich the experience of both animals and humans?

We conclude by noting two further and more pragmatic aspects with regards to which we should support human involvement concerning the care of animals and robots respectively.\\

\noindent
\textbf{Ensuring animal welfare.} While animal engagement to some extent can be provided through autonomous systems, ultimately every system depends on human involvement to take responsibility for monitoring, approving and overseeing its interactions. Responsibility for the welfare of the cats spread across multiple roles in Cat Royale: the operator and artist for immediate safety around the robot, and the animal welfare officer, owners and vet for general wellbeing. Similar functions will need to be supported for other systems that engage animals with robots, either through real-time monitoring and intervention or offline review and reconfiguration.

While during the design of Cat Royale we focused on the design of more than human-centred worlds~\cite{wakkary2021things} , we emphasised the cat's as primary non-human actors. While not focusing on every possible actor within a domestic ecology, we made deliberate choices to introduce cat grass and other, animal friendly, plants within the environment. This multispecies ecology could be understood to comprise an even broader flora \& fauna, emphasising the interaction with other non-human actors beyond the cats. While, during the analysis of Cat Royale, resulting in the findings presented in this paper, we focus on the cats as the sole non-human actor, other work has considered a range of other actors, including house plants and insects. One example is presented by, e.g.,~\cite{Bendel:2017:ladybird} who has investigated insect friendly robotic vacuum cleaners, in the context of the home, and the impact these have on the animals' welfare, which is typically disregarded. Further research, could expand this project's focus to encompass multispecies interactions within a broader ecological context in the home.\\


\noindent
\textbf{Robot wrangling and restoring order.} Some of the chaotic goings-on in Cat Royale that emerged from the cats' enthusiastic play with the robot reveal that humans will routinely need to intervene to find (e.g. Vignette~\ref{vig:manual_override}), gather and repair/replace objects; disentangle, reset, or re-calibrate the robot; and generally tidy up the mess that ensues animal-robot interactions. This is particularly the case as we cannot expect animals to do so for themselves---they will not judge or respect the limits of the robot, nor be able to fix any problems that occur. They are non-responsible users unconcerned for the robot as an actor in the system. While ideally the robot would be able to perform some of these tasks by itself (or perhaps other robots would), it seems that current robotic systems are nowhere near having the capabilities required to do so yet. The implication here is that everyday robot owners need to be enabled to wrangle the robot, for example by disentangling, resetting and re-calibrating it.

\subsection{Limitations}
The study focused on the design of a multispecies world with a bespoke enclosure for the interaction between three cats and a robotic arm. Behind the scenes, a control room with a variety of humans-in-the-loop (including artists, robot operator, and the animal welfare officer). However, while this robot world, or robot ecology, deliberately introduced other non-human actors, i.e., plants, into the ecology, the findings presented in this paper focus on the cats. Future domestic robots will encounter a wide variety of non-human actors including insects and plants. Future work might consider widening the meaning of the multispecies ecology view, and investigate how these worlds might be shaped to include a multitude of non-human actors.

Would a system like this be able to provide animal happiness in the wild, i.e., in an actual home? While Cat Royale probed questions related to autonomous systems, and investigated their capacity to take care and ultimately increase animal happiness, the project, by design, cannot provide results in relation to the system's capacity to replicate the observed effects in the wild. This investigation, informed by projects such as the one presented in this study, remains open for future work.
\section{Conclusion}
In this paper, we have presented the artist-led project Cat Royale, an exploratory investigation at the intersection between art and human/animal-robot interaction. 
We presented a detailed account on the design of the project, emphasising both the robot design and the design of the wider context in which it was situated, i.e., its 'robot world'. We also revealed that the importance of humans-in-the-loop cannot be understated. We described the unravelling of the events that transpired as part of the twelve day long installation, aiming to offer the reader insight into both the routine and the more challenging animal-robot interactions of Cat Royale. We then present two key implications which emerged through the Cat Royale project, which could inform future research and development of domestic robots. Firstly, we stress the importance of considering the environment beyond the robot design to include the design of the wider context, i.e., the physical world the robot and other actors are expected to share. Secondly, we highlight the need for designing multispecies interactions to include human involvement during robot supported enrichment for companion animals, as necessary functions implicated in animal-robot interaction (e.g., holding responsibility for animal welfare or recovering from breakdowns) cannot be outsourced to autonomous systems.

\begin{acks}
    This work was supported by the Engineering and Physical Sciences Research Council [grant number EP/V00784X/1] UKRI Trustworthy Autonomous Systems Hub.
\end{acks}

\bibliographystyle{ACM-Reference-Format}
\bibliography{biblio}
\end{document}